\documentclass{aastex62}
\definecolor{forestgreen}{rgb}{0.10, 0.50, 0.10}

\usepackage{amsmath}
\usepackage{mathtools}
\usepackage{footnote}
\usepackage{epstopdf}
\usepackage[T1]{fontenc}
\usepackage{hyperref}
\renewcommand{\textbf}[1]{{{#1}}}
\shorttitle{Fine structures of Solar S-bursts}
\shortauthors{Zhang et al.}

\begin{document}

\title{The frequency drift and fine structures of Solar S-bursts in the high frequency band of LOFAR}

\correspondingauthor{Pietro Zucca}
\email{zucca@astron.nl}
\author{PeiJin Zhang} 
\affil{CAS Key Laboratory of Geospace Environment,
	School of Earth and Space Sciences, \\
	University of Science and Technology of China,
	Hefei, Anhui 230026, China}
\affil{CAS  Center for the Excellence in Comparative Planetology,
    Hefei, Anhui 230026, China}
\affil{ASTRON, The Netherlands Institute for Radio Astronomy,\\
	 Oude Hoogeveensedijk 4, 7991 PD Dwingeloo, The Netherlands}
	 
\author{Pietro Zucca}
\affil{ASTRON, The Netherlands Institute for Radio Astronomy,\\
	 Oude Hoogeveensedijk 4, 7991 PD Dwingeloo, The Netherlands}

\author{ChuanBing Wang}
\affil{CAS Key Laboratory of Geospace Environment,
	School of Earth and Space Sciences, \\
	University of Science and Technology of China,
	Hefei, Anhui 230026, China}
\affil{CAS  Center for the Excellence in Comparative Planetology,
    Hefei, Anhui 230026, China}
\affil{Mengcheng National Geophysical Observatory, Hefei, Anhui 230026, China}

\author{Mario M. Bisi}
\affil{RAL Space, United Kingdom Research and Innovation (UKRI) - Science and Technology Facilities Council (STFC) - Rutherford Appleton Laboratory (RAL), Harwell Campus, Oxfordshire, OX11 0QX, U.K.}

\author{Bartosz Dąbrowski}
\affil{Space Radio-Diagnostics Research Centre, University of Warmia and Mazury in Olsztyn, Olsztyn, Poland}

\author{Richard A. Fallows}
\affil{ASTRON, The Netherlands Institute for Radio Astronomy,\\
	 Oude Hoogeveensedijk 4, 7991 PD Dwingeloo, The Netherlands}
	 
\author{Andrzej Krankowski}
\affil{Space Radio-Diagnostics Research Centre, University of Warmia and Mazury in Olsztyn, Olsztyn, Poland}

\author{Jasmina Magdalenic}
\affil{ Solar-Terrestrial Centre of Excellence—SIDC, Royal Observatory of Belgium, 1180 Brussels, Belgium}

\author{Gottfried Mann}
\affil{Leibniz-Institut für Astrophysik Potsdam (AIP), An der Sternwarte 16, D-14482 Potsdam}

\author{Diana E. Morosan}
\affil{Department of Physics, University of Helsinki, P.O. Box 64, FI-00014 Helsinki, Finland}

\author{Christian Vocks}
\affil{Leibniz-Institut für Astrophysik Potsdam (AIP), An der Sternwarte 16, D-14482 Potsdam}

\begin{abstract}
Solar S-bursts are short duration ($<1$ s at decameter wavelengths) radio bursts that have been observed during periods of moderate solar activity, where S stands for short. {The frequency drift of S-bursts can reflect the density variation and the motion state of the electron beams. In this work, we investigate the frequency drift and the fine structure of the S-bursts with the LOw Frequency ARray (LOFAR). We find that the average frequency drift rate of the S-bursts within 110\,--\,180\,MHz could be described by $df/dt=-0.0077f^{1.59}$.} With the high time and frequency resolution of LOFAR, we can resolve the fine structures of the observed solar S-bursts. A fine drift variation pattern was found in the structure of S-bursts (referred to as solar Sb-bursts in this paper) during the type-III storm on 2019 April 13, in the frequency band of 120\,--\,240 MHz. The Sb-bursts have a quasi-periodic segmented pattern, and the relative flux intensity tends to be large when the frequency drift rate is relatively large. This kind of structure exists in about 20\% of the solar S-burst events within the observed frequency range. We propose that the fine structure is due to the density fluctuations of the background coronal density. We performed a simulation based on this theory which can reproduce the shape and relative flux intensity of the Sb-bursts. This work shows that the fine structure of solar radio bursts can be used to diagnose the coronal plasma.

\end{abstract}



\section{Introduction}
\label{sec:intro}

Solar S-bursts are short duration radio bursts with narrow bandwidths. They were first identified by \cite{ellis1969fine} as a new type of short duration radio bursts, and named 'fast drift storm bursts'. Later, \cite{mcconnell1982spectral} carried out a statistical study of these 'fast drift storm bursts' and renamed them to solar S-bursts as they showed similar characteristics to Jovian S-bursts. The statistical results \citep{mcconnell1982spectral, melnik2010solar, morosan2015lofar, morosan2018characteristics, clarke2019properties} showed that the {FWHM} duration the S-burst is about 50ms at a given frequency, the {instantaneous} bandwidth is about 120\,kHz. Solar S-bursts always occur superimposed on a background of other solar radio activity, such as type IIIs, type IIIbs and spikes. The frequency drift rate of S-bursts is generally about 1/3 to 1/2 of the accompanying type III bursts. It is also found that the instantaneous frequency width of solar S-bursts increases linearly with frequency \citep{melnik2010solar, clarke2019properties}. \cite{melrose1982fine} argued that solar S-bursts are a similar phenomenon to the drift-pair bursts with reflections. A recent observation of drift pairs bursts  indicates that the drift pairs have almost the same frequency drift rate as the S-bursts \citep{stanislavsky2017solar}. \cite{mcconnell1981fine} identified the  fringed structure in type-S burst, which was reported to exist in 1\% of the type-S events in the frequency band of 30\,--\,82 MHz. The fringed structure is regularly spaced at about 100\,kHz between the bright stripes. The single elements of the fringes have narrow band width down to 10\,--\,15 kHz \citep{mcconnell1982spectral, mcconnell1983evidence}. So far, there have been very few studies on the fine structure of solar S-bursts because of the limited sensitivity and resolution of previous observations.

LOFAR is an advanced radio antenna array \citep{van2013lofar}. It has two types of arrays observing in two frequency bands, the Low Band Array (LBA) in the frequency range of 10\,--\,90 MHz and High Band Array (HBA) in the frequency range of 110\,--\,250 MHz. LOFAR has 51 stations, 38 of them located in the North-East of the Netherlands, and 13 international stations located in Germany, Poland, France, Sweden, Ireland, and UK. LOFAR is capable of a variety of processing operations including correlation for standard interferometric imaging, the tied-array beam-forming. With the high time and frequency resolution of LOFAR observations, we can resolve the fine structure of solar S-bursts.

The emission mechanism of S-bursts is still an open question. Several mechanisms could be used for the interpretation of S-bursts. For example the electron cyclotron maser (ECM) emission was first proposed by \cite{wu1979theory} for the auroral kilo-metric radiation (AKR) of the Earth, and used to interpret the fine structure of type III radio burst (the type IIIb burst) \citep{wang2015scenario}. The ECM wave is generated in the condition of $f_{ce}/f_{pe} > 1$, where $f_{ce}$ is the electron cyclotron frequency and $f_{pe}$ is the local plasma frequency. The condition of $f_{ce}/ f_{pe} > 1$ requires tenuous plasma and relatively strong magnetic fields. \cite{melnik2010solar} proposed a model for S-bursts, in which the electromagnetic wave is generated in the processes of the coalescence of fast magneto-sonic waves with Langmuir waves. According to a recent study by \citet{morosan2016conditions}, the most likely emission mechanism for meter/decameter frequency radio bursts is plasma emission, since the magnetic field at the height of these radio bursts is not high enough to produce ECM emission. However, \cite{kolotkov2018origin} argued that the magnetic field would be sufficient to excite ECM emission when there is enough complexity in the magnetic field configuration. In addition, the condition of the ECM can be satisfied with the assumption of density-depleted magnetic flux tubes \citep{wu2002generation,wang2015scenario}. Based on this assumption, the true source region, where the radio wave is generated by ECM, is inside the tube, and the wave cannot leave the tube until it arrives at an altitude where the local exterior cut-off frequency is equal to the wave frequency (the apparent source region). Which means, the apparent source of ECM emission also satisfies that the wave frequency is approximately equal to the local plasma frequency of 'observed source region'. Although the emission mechanism of solar S-bursts is still debated, the observed emission originates in regions of local plasma frequency levels \citep{mcconnell1983evidence,melnik2010solar,morosan2015lofar}. Thus, we assume that S-bursts are generated close to the plasma frequency.

The coronal plasma has various length scale of density fluctuations from up to a few solar radii and down to kilometer scale \citep{woo1996kilometre}. The density fluctuations of the the corona plasma can result in the fine-structured solar radio bursts, and these fine structures can be used to study the density fluctuations of the corona. \cite{kolotkov2018origin} used the quasi-periodic drifting fine structures of type III radio bursts to study the density variation caused by fast magneto-acoustic waves. \cite{loi2014production} used simulations to show that the fine structure in type III radio bursts could be due to the background density fluctuations. \cite{mugundhan2017solar} and \cite{Chen_2018} used the type IIIb radio bursts as a tracer to find that the spectrum of the density variation derived from the frequency gap between the striae fits well to the one-dimensional Kolmogorov spectral index (5/3) \citep{frisch1995turbulence}. \cite{mann1989interpretation} used the perturbed density to interpret the fine structures in type IV radio bursts. In this work, we inspected the fine structure of S-bursts (referred to as Sb-bursts in the following), and demonstrated that the fine structure of Sb-bursts can be used to inspect the small scale (mega-meter) density fluctuations.

This paper is arranged as follows, in Section 2, we show {observation results on the frequency drift rate of S-bursts from LOFAR HBA observations,} the dynamic spectrum of six Sb-bursts, and the anaysis of their characteristics. In Section 3, we qualitatively show that the relationship between the frequency drift rate and the relative flux intensity in the dynamic spectrum can be interpreted by Coulomb absorption of the wave with electron density fluctuations. Then, we show the simulation results which can reproduce the shape and relative flux intensity of the Sb-bursts. In Section 4 we present our conclusion and discussion.

\begin{figure}[h]
	\centering
	\includegraphics[width=0.9\linewidth]{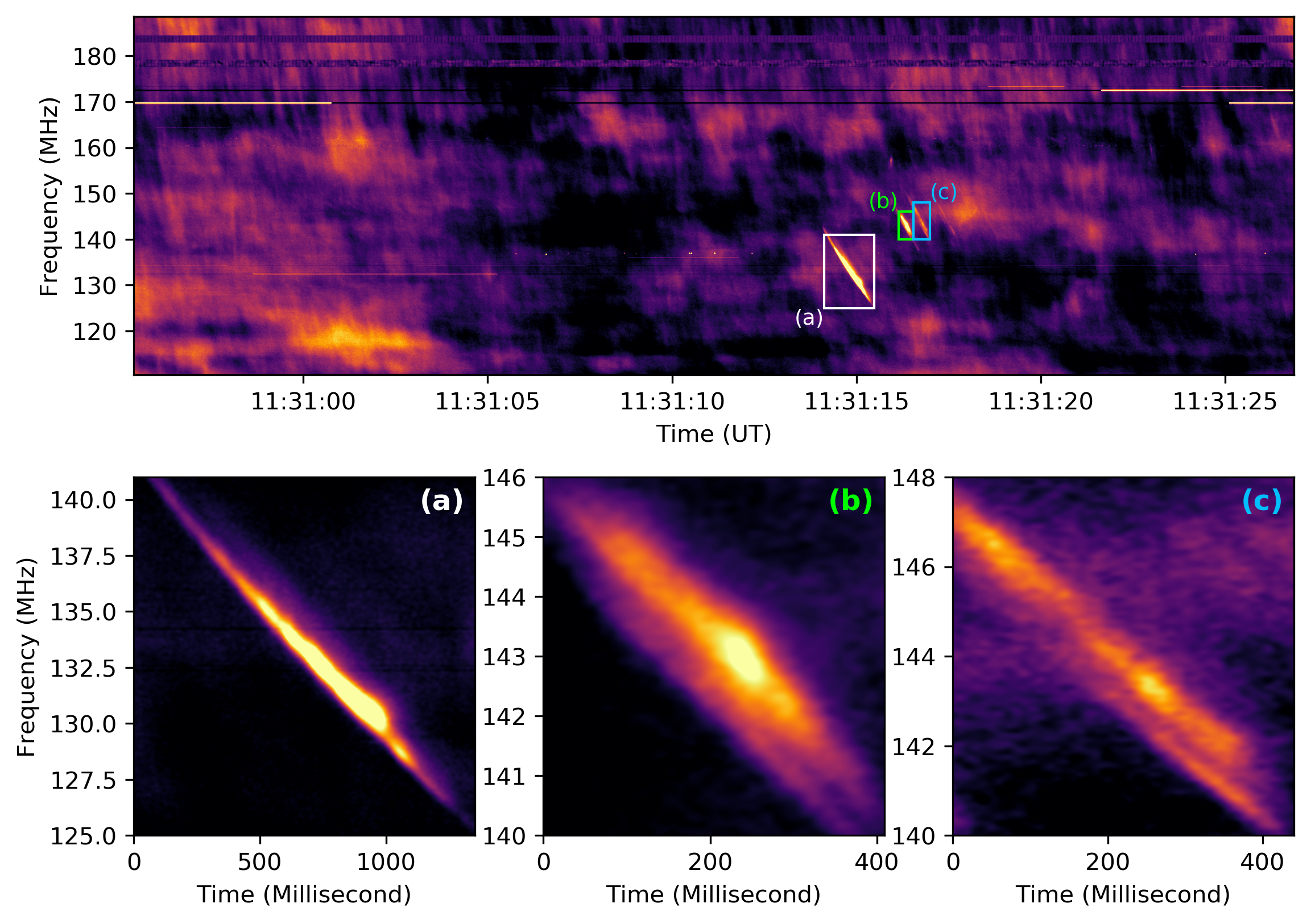}
	\caption{An example of the dynamic spectrum of HBA during the type-III storm on 2019 April 13. Three S-burst is labeled as (a,b,c) in the upper panel, the zoom-in view of these there S-burst is shown in the lower panel.}
	\label{fig:overview}
\end{figure}

\section{S-bursts Observations}

In this study, we used dynamic spectra observations of the Sun obtained by a single HBA beam produced using the LOFAR core stations \citep{van2013lofar}. The synthesized beam which points at the solar center has a main lobe width (about 0.1 Degree) smaller than the angular width of the Sun, which can help us avoid the emission from other directions and make sure the radio bursts originate from the Sun. The beam-formed observations provide a high-time and frequency resolution of 10.5\,ms and 12.2\,kHz. The observations were taken in the time interval from 05:40:00\,UT to 17:39:59\,UT on 2019 April 13. In this time interval we also observed a type III storm. An example of a dynamic spectrum during this time is shown in Figure \ref{fig:overview}. The S-burst shown in Figure \ref{fig:overview}(a) is a typical S-burst, while the S-burst in Figure \ref{fig:overview}(b) and (c) shows multi-lanes, similar to the "drift pairs" bursts. {The band-splitting width of the observed drift pair is about 0.5\,MHz, the time difference between the lanes of the drift pairs is about 50 millisecond.}

\subsection{Frequency Drift Rate of S-Bursts}

\begin{figure}[ht]
	\centering
	\includegraphics[width=0.5\linewidth]{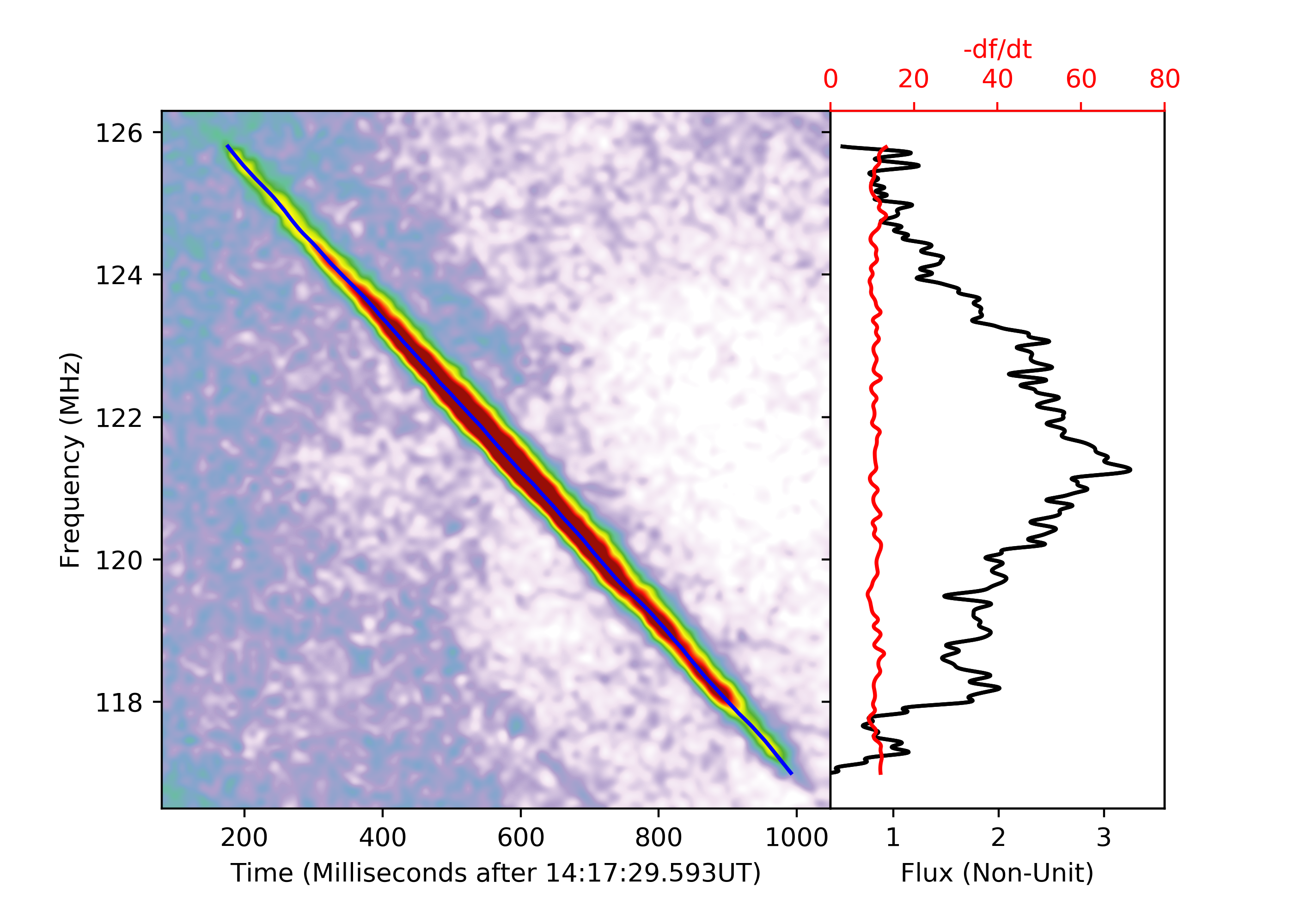}
	\caption{The dynamic spectrum shows an isolated S-burst. The right sub-panel shows the frequency drift rate (red line), and the relative flux intensity of this S-burst (black line).}
	\label{fig:2}
\end{figure}

We find 204 S-bursts during our 12 hour observation of the Sun. Figure \ref{fig:2} shows the dynamic spectrum of a typical S-burst in HBA, and its frequency drift rate and relative flux intensity. In the left sub-panel of Figure \ref{fig:2}, the blue line represents the center line of the S-burst, namely the frequency drift line. The frequency drift line is obtained using an iteration algorithm called ACBone \citep{zhang2018type}. This algorithm was used previously to statistically study the frequency drift rate of type III bursts. {The code of ACBone is available online at \href{https://github.com/Pjer-zhang/TypeIIIRadioBurstRecognition}{Github}}. From the frequency drift line, we can obtain the frequency drift rate and the flux intensity of the S-burst at different frequencies, which is shown in the the right sub-panel of Figure \ref{fig:2} with red and black curves, respectively. For the S-bursts shown in Figure \ref{fig:2} we can see that the frequency drift rate is stable near 12\,MHz/s, corresponding to the smooth frequency drift line in dynamic spectrum. The intensity shows a trend of increase and decrease at the first and second half of this event.
 
\begin{figure}[t]
	\centering
	\includegraphics[width=0.5\linewidth]{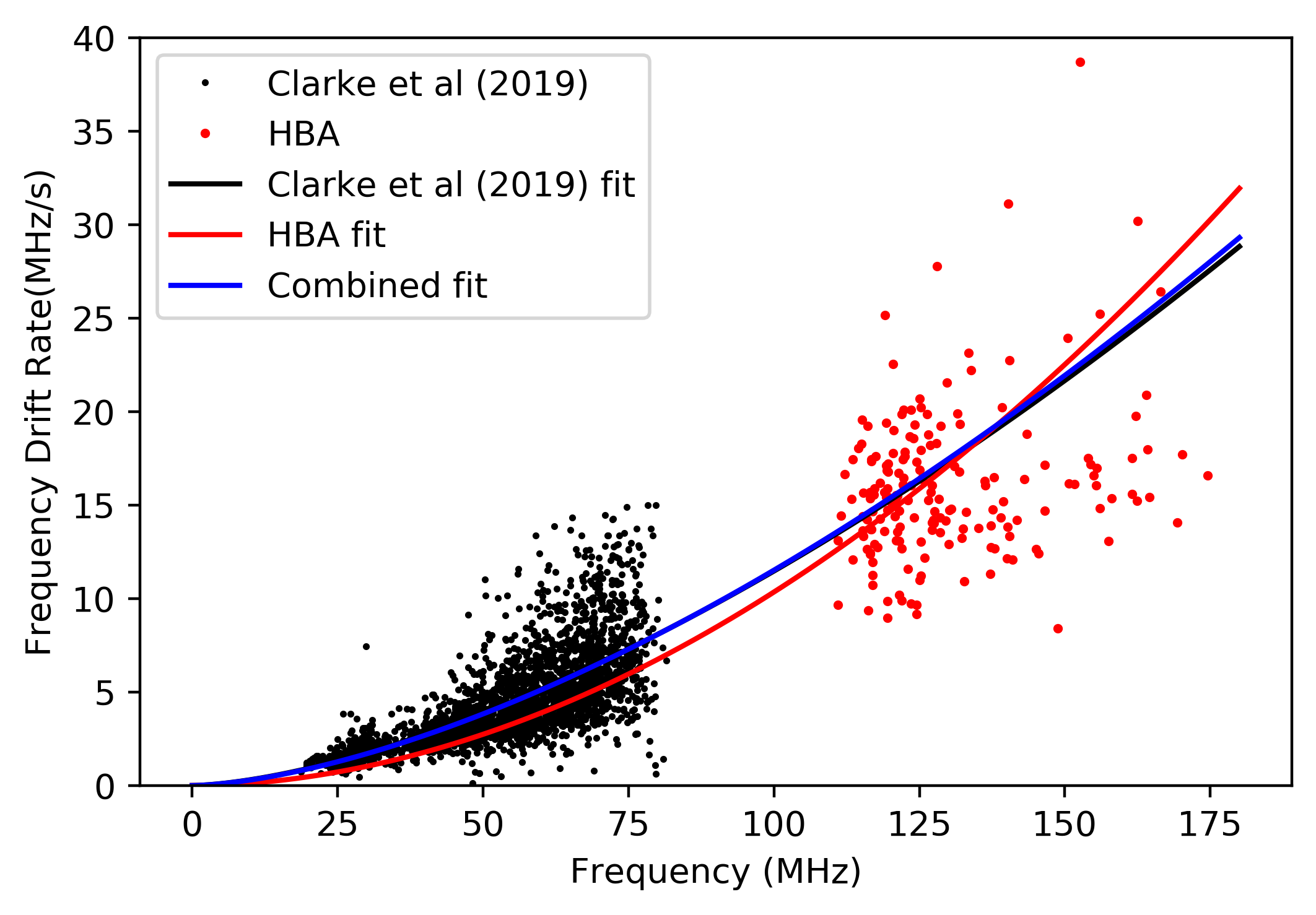}
	\caption{The frequency drift rate of S-bursts. The black points contains the previous result of low frequency band of 20\,MHz to 70\,MHz including observation from \cite{mcconnell1982spectral,dorovskyy2017properties,morosan2018characteristics,clarke2019properties}. The red points mark the 204 events in the frequency range of 120\,--\,180\,MHz on 2019 April 13. \ {The red line shows the fitting result to the events in HBA ($df/dt= -0.0015 f^{1.92}$), the black line shows the fitting result in low frequency ($df/dt= -0.0084 f^{1.57}$), the blue lines  shows the combined fitting result of the S-bursts in HBA and low frequency ($df/dt= -0.0077 f^{1.59}$).}}
	\label{fig:3}
\end{figure}

For each event, we used a linear fit to the frequency drift line to get an average frequency drift rate ($df/dt$). The frequency drift rate of all S-bursts is shown in Figure \ref{fig:3}, along with the previous results in low frequency. From Figure \ref{fig:3} we can see that the frequency drift rate of S-bursts agrees well with the fit of the low frequency observations (blue line in Figure \ref{fig:3}). \ {We fit the frequency drift line of these S-bursts in HBA to the power-law equation ${df}/{dt} = -a f^b $, the fitting result is $df/dt= -0.0015 f^{1.92}$. The combined fitting result of the S-bursts in HBA and the S-bursts in low frequency is $df/dt= -0.0077 f^{1.59}$. The relation between the frequency and frequency drift rate is similar in low frequency range (10\,--\,80 MHz) and high frequency range (120\,--\,180 MHz), as shown in Figure \ref{fig:3}. }

\subsection{Fine Structure of Sb-Bursts}
\begin{figure}[h]
	\centering
	\includegraphics[width=0.48\textwidth]{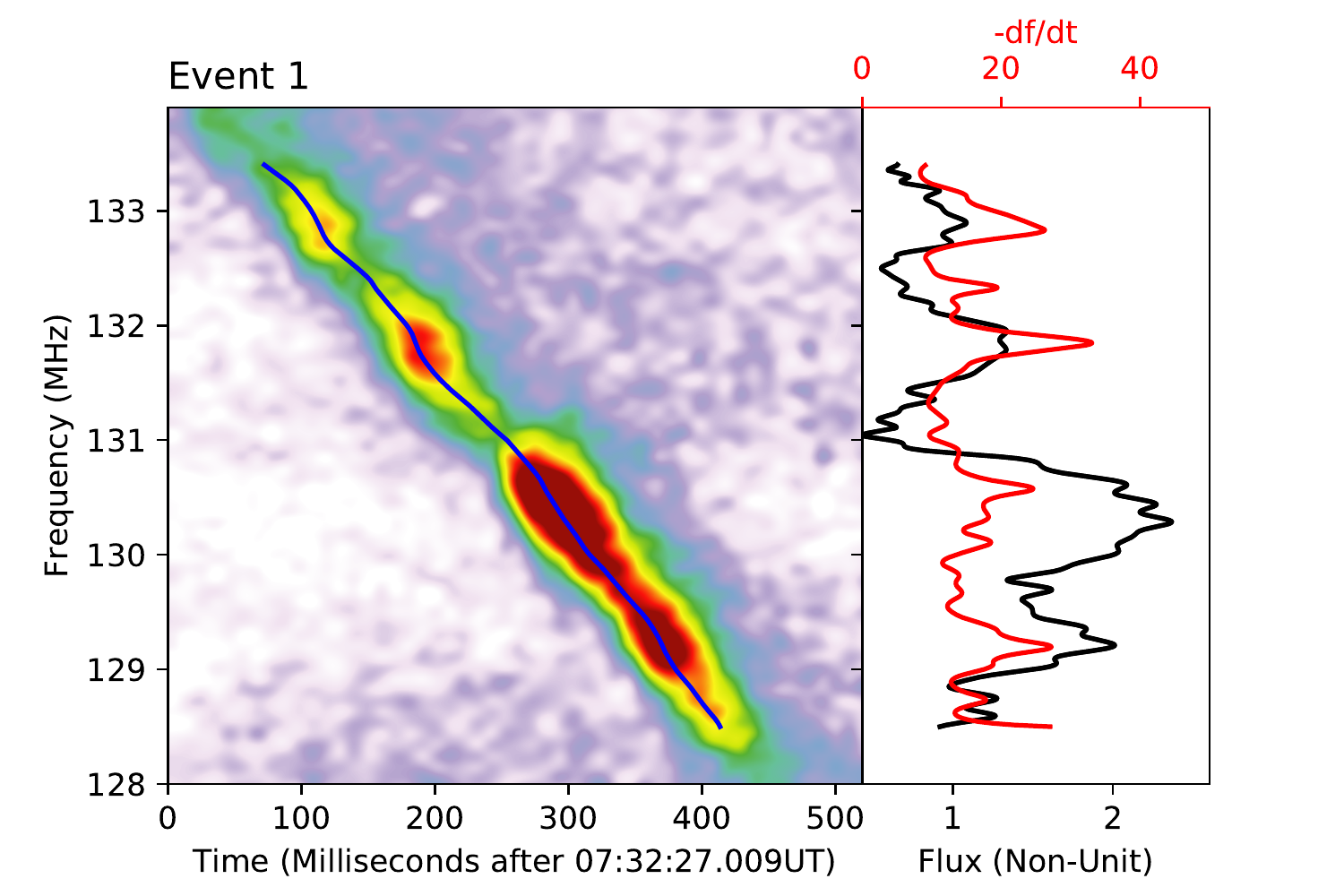}
	\includegraphics[width=0.48\textwidth]{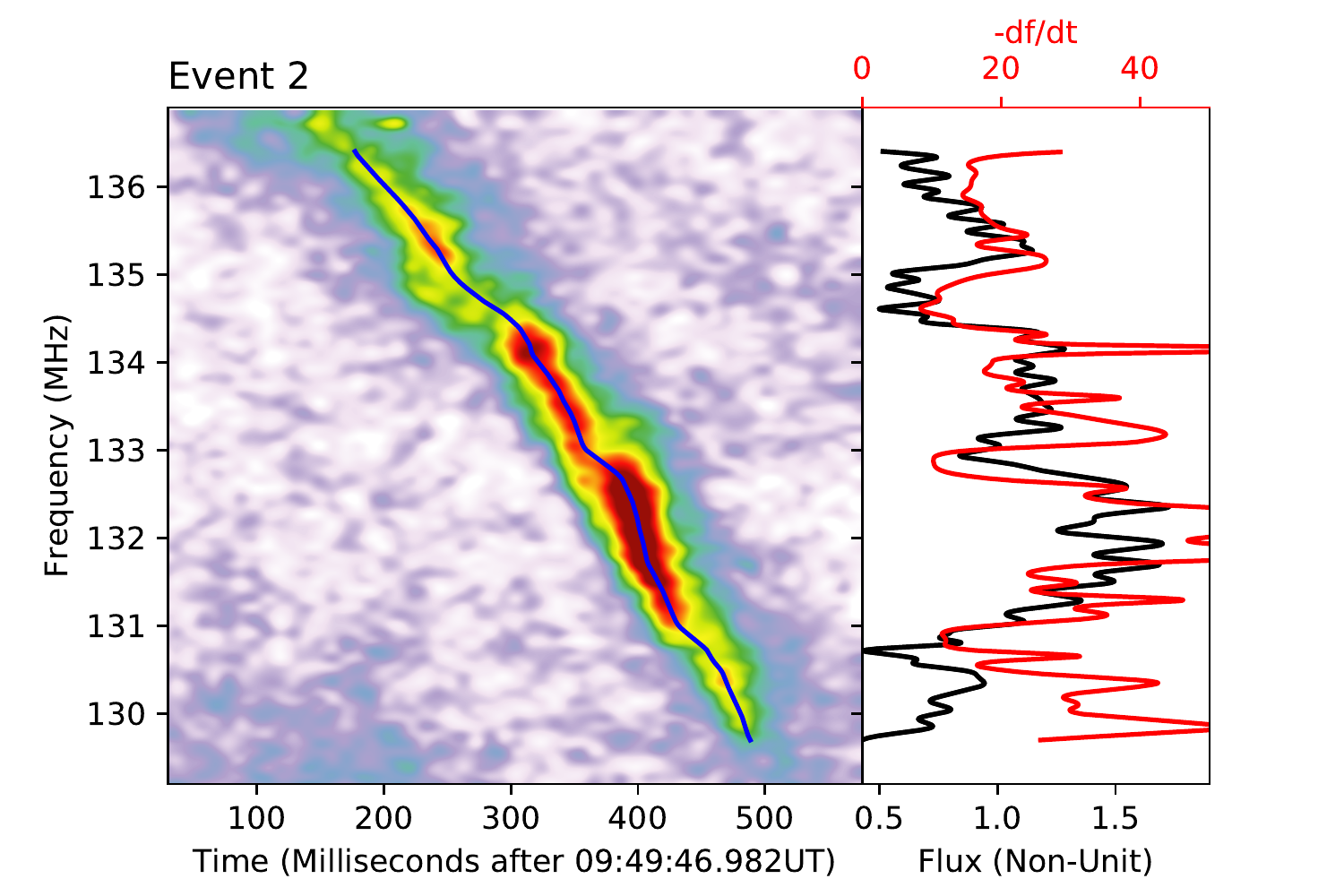}
	
	\includegraphics[width=0.48\textwidth]{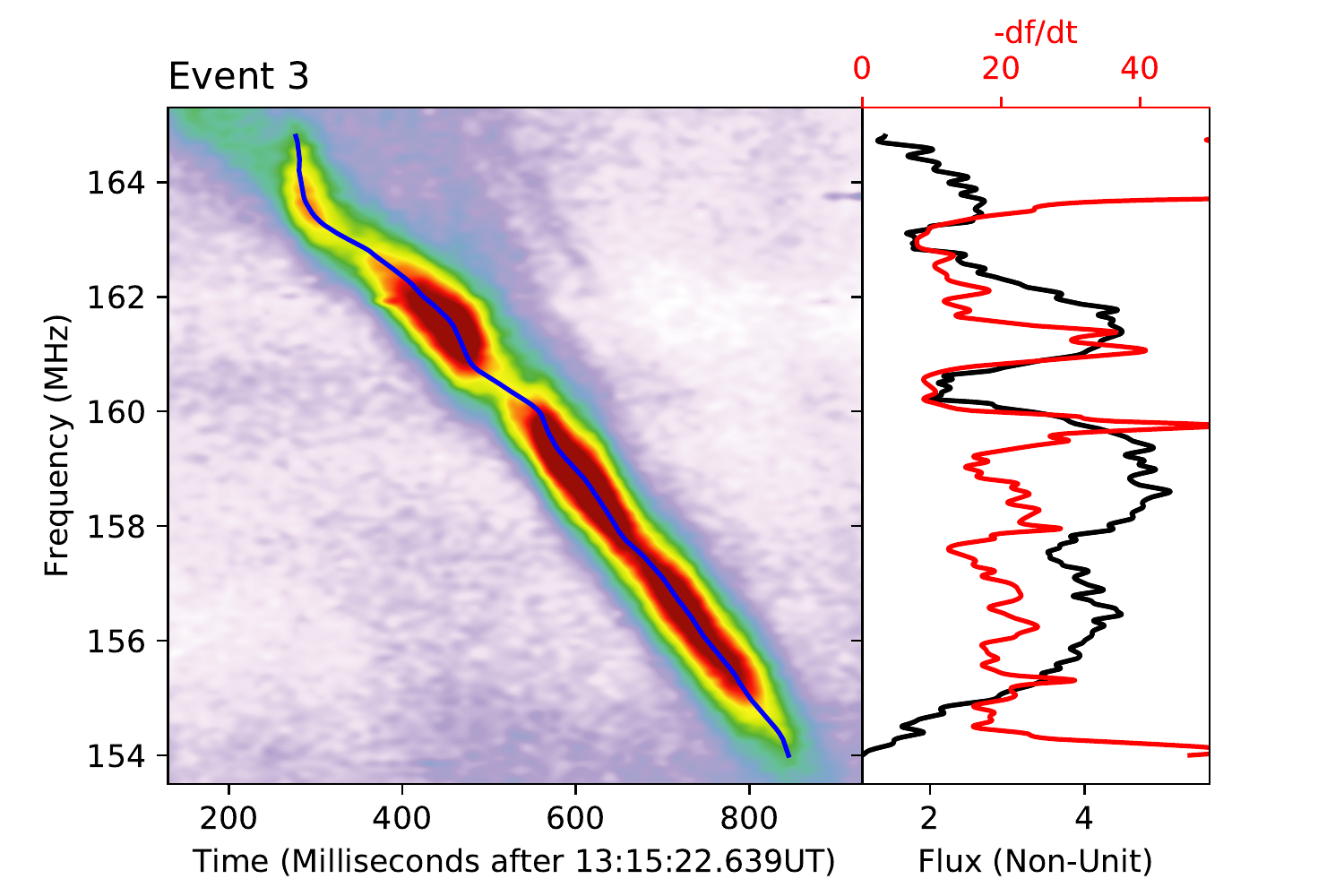}
	\includegraphics[width=0.48\textwidth]{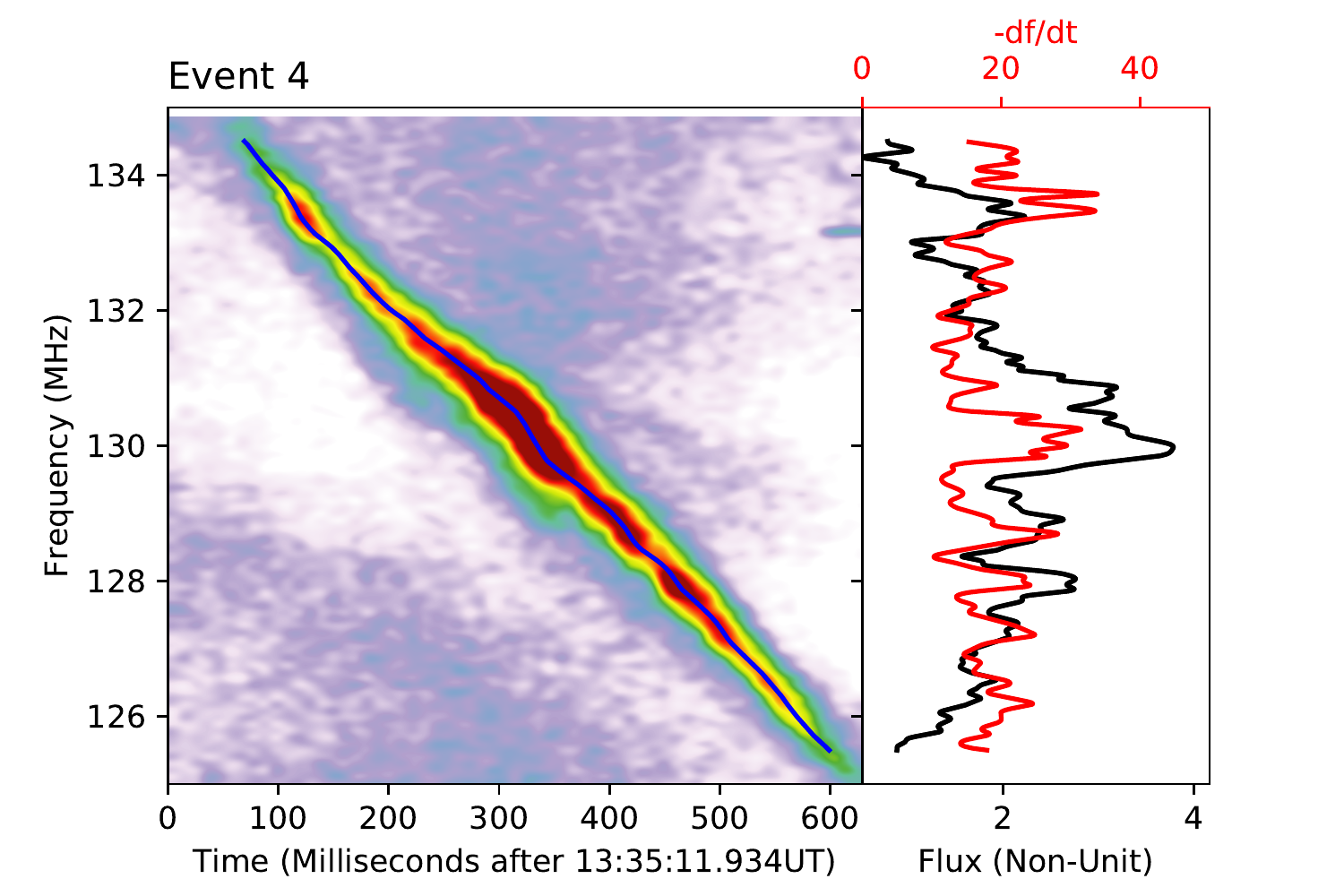}
	
	\includegraphics[width=0.48\textwidth]{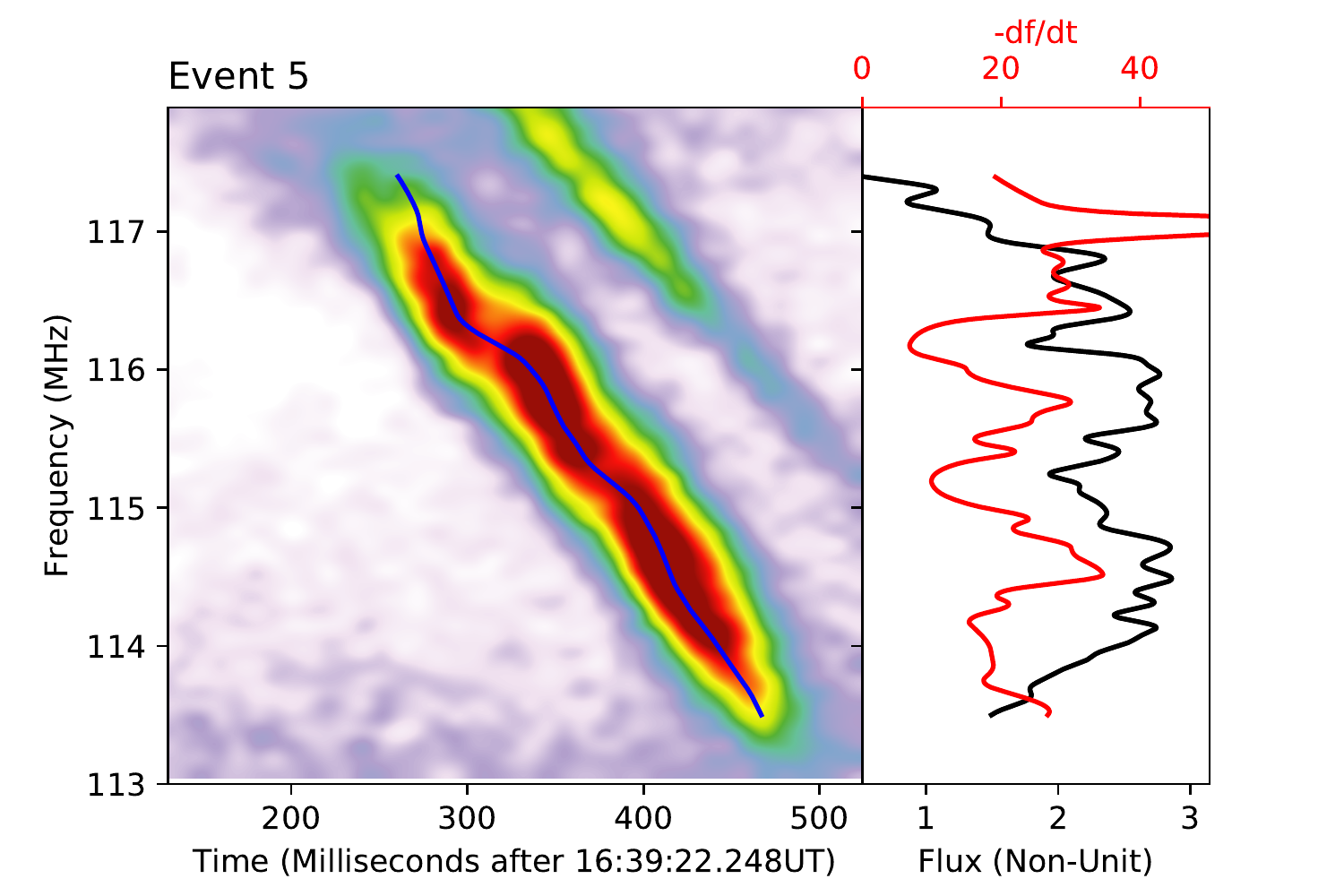}
	\includegraphics[width=0.48\textwidth]{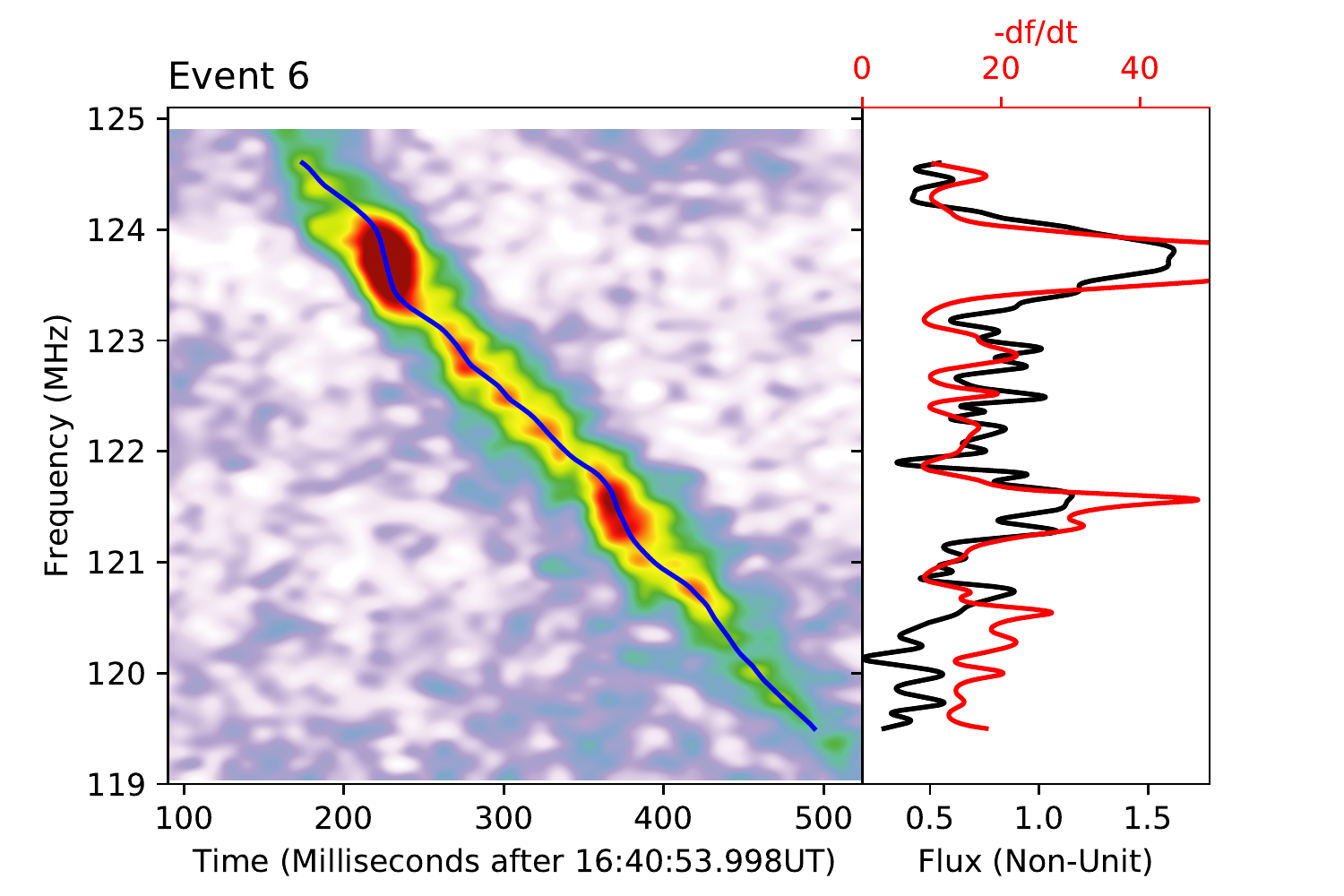}
	\caption{
		The dynamic spectrum, the frequency drift rate, and the relative flux intensity of six Sb-bursts with fine structures on 2019 April 13. In each panel, the left sub-panel shows the dynamic spectrum with a frequency drift line, taken as the center-line of the burst. The right sub-panel shows the frequency drift rate (red line) and the relative flux intensity (black line) of the S-bursts.  The left and right sub-panel shares the same y-axis.
		\label{fig:1}}
\end{figure}

Out of these 204 events, we find that {some of the S-bursts have a discontinuous structure}. As examples, Figure \ref{fig:1} shows the dynamic spectrum, the frequency drift rate, and the relative flux {intensity} of six Sb-bursts with fine structures.

\begin{figure}[h]
	\centering
	\includegraphics[width=0.53\textwidth]{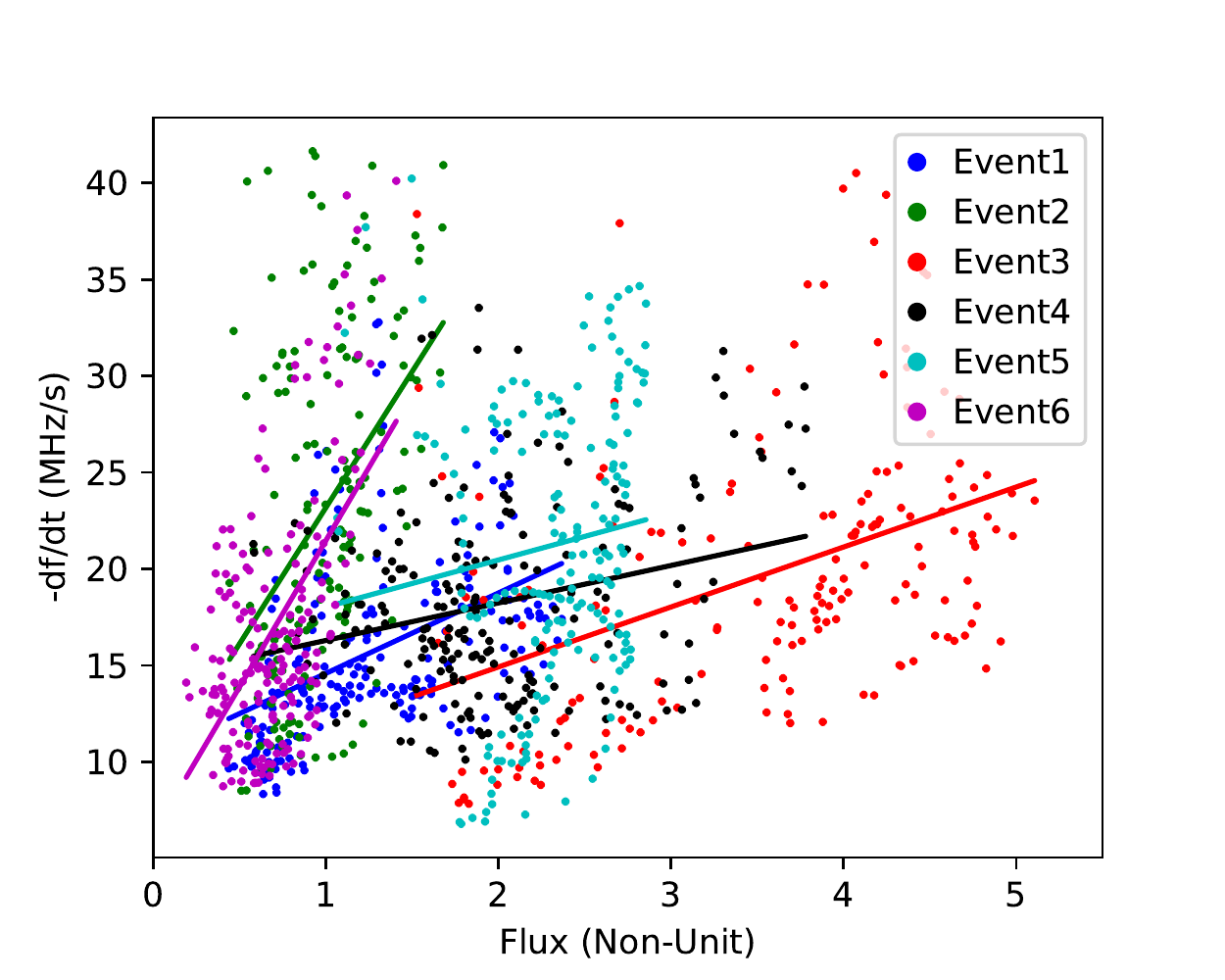}
		\caption{
		The linear fit of the frequency drift rate and flux intensity. Six colors marks the six events in Figure \ref{fig:1}, the data points are obtained along the frequency drift line of each event, the solid line is the linear fit of the data points. \label{fig:fit}}
\end{figure}

Figure \ref{fig:1} shows an interesting pattern in the dynamic spectrum of the Sb-bursts. The Sb-bursts are segmented like chains, and the frequency drift line is relatively steeper near the local maximum of each segment. The frequency drift rate ($-df/dt$) and the the flux intensity ($I_{obs}$) have a similar trend, namely the flux intensity tends to be large when the frequency drift rate is large. \ {This trend can be seen in every left sub-panel of Figure \ref{fig:1}}. 

\ {For these six events, we did a Pearson test and a linear fit between $df/dt$ and $I_{obs}$, and the results are shown in Figure \ref{fig:fit} and Table \ref{tab:0}. The Pearson correlation coefficient ($P_{coef}$) and the linear fit slope of these six selected events are all positive, which indicates positive correlation between the frequency drift rate and the intensity. Among these six events, Event 6 has the largest correlation coefficient. The $df/dt$ and $I_{obs}$ of Event 1,2, and 3 have significant correlation, Event 4 has moderate positive correlation, while Event 5 has weak correlation. In the following, we will propose a model to explain the positive correlation between the frequency drift rate and the flux intensity in the discontinuous structure of Sb-bursts.}

\begin{table}[h]
	\centering
	\begin{tabular}{ccc}
		\hline
		& Pearson r & Slope \\
		\hline\hline
		Event 1 & 0.44      & 4.16  \\
		Event 2 & 0.45      & 14.10 \\
		Event 3 & 0.42      & 3.11  \\
		Event 4 & 0.25      & 1.94  \\
		Event 5 & 0.12      & 2.44  \\
		Event 6 & 0.56      & 15.17 \\
		\hline
	\end{tabular}
\caption{The Pearson correlation coefficient and slope of the linear fit between $-df/dt$ and $I_{obs}$. \label{tab:0}}
\end{table}


\section{Generation of the Fine Structure}

In this section, we attempt to explain the shape and flux intensity in the fine structure of the Sb-burst by assuming the electron beams pass through a plasma with fluctuating background electron densities.

\subsection{The Relationship between Optical Depth and Frequency Drift Rate}

When the electron beam is propagating outward at a speed of $v_b$, the frequency drift rate can be expressed as 
\begin{equation}
\frac{df}{dt} = v_b \frac{df_{pe}}{d h } \propto  \frac{df_{pe}}{d h}.  \label{eq:dfdt} \quad
\end{equation}
The frequency drift rate is in proportional to the \ {decreasing rate of the local plasma frequency with altitude}. According to the plasma emission theory, the wave is generated near the plasma frequency $f_{pe}$, which can be expressed by $f_{pe} [\rm{Hz}] = 8.98\times 10^{3} \sqrt{N_e [\rm{cm^{-3}}]}$. Here, $N_e$ is the electron number density.  Thus, the increasing and decreasing trend in the frequency drift rate in Figure \ref{fig:1} can be related directly to fluctuations in the background electron density.

The variation in the flux intensity can be interpreted by {Coulomb collision absorption of the wave}. The attenuation caused by collisional absorption is given by \cite{benz2012plasma} (p265, Eq. 11.2.2):

\begin{align}
	I_{obs} &=I_0 e^{-\tau_c} \label{eq:0}\\
\tau_c &= \int_{s_0}^{\inf} \kappa(s) ds \label{eq:1}\\
\kappa(s) &\approx 9.88\times 10^{-3} \frac{N_e\sum_i Z_i^2N_i  \ln \Lambda_t}{f_s^2T^{3/2} \mu_r}  [\textrm{cm}^{-1}], \nonumber \quad
\end{align}
where $\tau_c$ is the Coulomb collisional optical depth, $\kappa(s)$ is the collisional damping rate of the radio wave, $s $ [cm] is the distance along the ray path, and $\mu_r$ is the refraction index. For a fully ionized plasma with solar abundances $\sum_i Z_i^2N_i = 1.16N_e$. $\ln\Lambda_t$ is the Gaunt factor. ${T}$ [K] is the thermal temperature. $f_s$  $ (\gtrsim f_{pe})$ is the frequency of the wave from the source in units of Hz. 

Here we consider that the wave generated in the corona is propagating upwards, and for simplicity, we also use one-dimensional (1-D) wave propagation. The refraction index in the plasma is
\begin{equation}
\mu_r = \frac{v_g}{c} = \sqrt{1-\frac{f_{pe}^2}{f_s^2}} \quad,
\label{eq:speed}
\end{equation}
where $v_g$ is the wave group speed, $c$ is the speed of light in vacuum. Thus, for the wave of a given frequency $f_s$, which is excited at the height $h_0$ where $f_s\approx f_{pe}(h_0)$, the collisional optical depth can be rewritten as:
\begin{align}
	\tau_c &= \int_{h_0}^{inf} \kappa_0 f_{s}^2 \frac{f_{pe}^4}{f_s^4} \frac{1}{\sqrt{1-\frac{f_{pe}^2}{f_s^2}}} dh \quad,
	\label{eq:main}
\end{align}
where $\kappa_0$ is a constant expressed as
\begin{equation}
	\kappa_0 = 1.76\times10^{-18} \left(\frac{\ln\Lambda_t}{T^{3/2}}\right) [\rm{cm^{-1} Hz^{-2}}] \quad.
\end{equation}
{Though the integrand in 
Equation \ref{eq:main} is singular at $f_{pe} = f_s$, and the integration can be performed. In addition, the integral decreases quickly with decrease of the frequency ratio $f_{pe}/f_s$. This indicates that the main contribution of $\tau_c$ comes from the plasma near the radio source region with $f_{pe}\lesssim f_s$, since the corona electron density is generally decreasing with the increase of altitude. As approved in Appendix \ref{ap:1}, the optical depth \textbf{is determined mainly by the function of $f_{pe}$ near the emitting region and} is inversely proportional to the local decreasing rate of the background plasma frequency near the radio source region, namely,
\begin{equation}
\tau_c (f_s) \propto \left(-\left.{\frac{df_{pe}}{dh}}\right|_{f_{pe}=f_{s}}\right)^{-1}.
\label{eq:tau}
\end{equation}}

{Equation \ref{eq:tau} indicates that}, when {the decreasing rate of plasma frequency locally (or the frequency drift rate according to Equation \ref{eq:dfdt})} is large, the optical depth of the corresponding radio wave is relatively small, and the wave is less absorbed. {If there is wavelike variation of the local corona density with altitude, one would observe similar variation of the frequency drift rate and flux intensity with the radio frequency.} Therefore, Equation \ref{eq:dfdt} and \ref{eq:tau} can qualitatively explain the reason why the segments with larger absolute value of frequency drift rate tend to be brighter in the dynamic spectrum. {While some assumptions were made for the derivation of Equation \ref{eq:tau}, it can qualitatively show the trend. In the following subsection, we present the numerical integration results based on a corona density model with fluctuation.}

\subsection{Numerical Calculation}
{For a given background electron density distribution $N_e(h)$, the frequency drift line $f(t)$ and the relative flux intensity of the wave can be obtained by numerically integrating Equation \ref{eq:dfdt} and Equation \ref{eq:main}, respectively.} In the calculation we made several assumptions:

\begin{itemize}
	\item We assume 1-D wave propagation along the line of sight.
	\item The S-burst is generated by an electron beam propagating outward in the corona, exciting the wave at the local plasma frequency ($f_{pe}$), and the wave starts to propagate along the radial direction.
	\item {The corona electron density can be described as
	\begin{equation}
	    N_e(r) = N_{e0}(r) + \Delta N_e(r),
	\end{equation}
	where the background density $N_{e0}$ is chosen to be four times the density model of  \cite{saito1977study},
	\begin{equation}	 N_{e0}(r) = 4\left(\frac{1.36\times10^6}{r^{2.14}} + \frac{1.68\times10^8}{r^{6.13}}\right) [\textrm{cm}^{-3}].
	\nonumber
	\end{equation}
	Here the heliocentric distance $r$ is in the unit of solar radius. The disturbed electron density is expressed as
  $$\Delta N_e(r) = \Delta N_{e0} \sin(2\pi r/\lambda ), $$
  where $\Delta N_{e0}=2.85\times 10^5 \ \rm{cm^{-3}}$ is the amplitude of the density fluctuation,  $2\pi/\lambda=3000$ corresponds to a wavelength of about 1.5\,Mm.}
	\item The intensity of the wave generated by the beam does not vary drastically during the burst.
	\item The wave is absorbed by Coulomb collisions during its propagation according to Equation \ref{eq:0}. 
\end{itemize}

\begin{figure}
	\centering
	\includegraphics[width=0.6\linewidth]{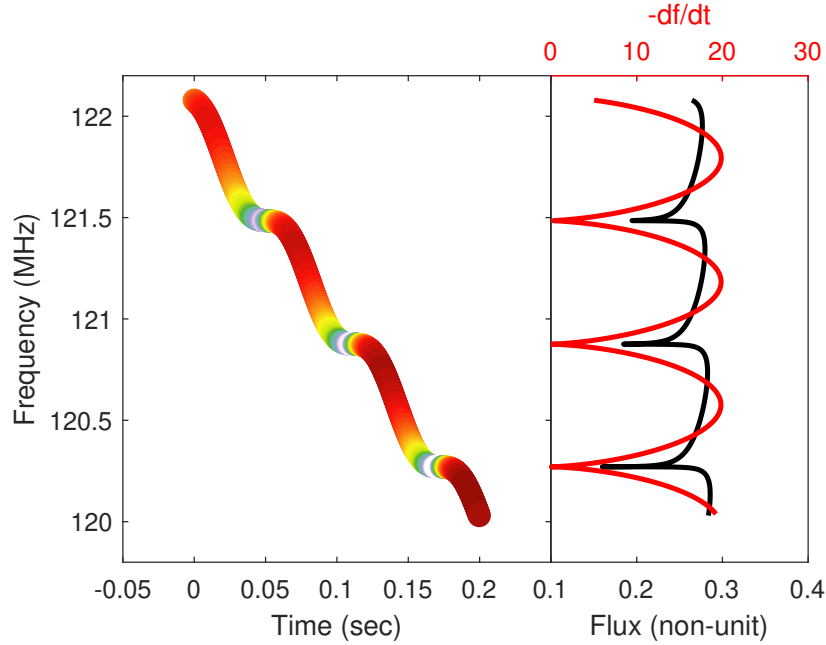}
	\caption{The simulated frequency drift line and the flux intensity, the color in the left sub-panel indicates the relative intensity. In the right sub-panel, the red line shows the frequency drift rate of this Sb-burst for different frequency, the black line shows the relative intensity calculated with Equation \ref{eq:0} and \ref{eq:main}. {The electron beam speed is set to 0.08 $c$ in the simulation.}}
	\label{fig:simulation}
\end{figure}

{With these assumptions,} when the electron beam passes through the perturbed plasma, it will experience {quasi-periodic} changes in the density gradient, which corresponds to the wavy structure in the frequency drift of the burst. {We calculated the optical depth $\tau_c$ between the generation site and the observer for the wave of different frequencies using Equation \ref{eq:main}. The integration range is from the position with local plasma frequency of $f_{pe} = (1-10^{-7})f_s$ to 200 solar radii.} The relative flux intensity is calculated using Equation \ref{eq:0}, then we overlap the relative intensity to the frequency drift line to simulate the dynamic spectrum of an S-burst with fine structure. The simulated dynamic spectrum is shown in Figure \ref{fig:simulation}, from which we can see that it can reproduce the wavy frequency drift line and the segmented intensity distribution \textbf{similar to our observation. The numerical calculations \footnote{The source code is available at a online repository( \url{https://github.com/Pjer-zhang/SBurst-Numerical})} is performed using the integral procedure in \href{https://www.mathworks.com/help/matlab/ref/integral.html}{MATLAB}, where the various step lengths are arranged by the procedure to achieve the optimal precision \citep{shampine2008vectorized}}.

\section{Conclusion and Discussion}

In this work, we find 204 S-bursts within the 12 hours observations in the frequency range of 120\,--180\,MHz of LOFAR on 2019 April 13. The frequency drift rate of the events are in good accordance with the previous results obtained in low frequency range \citep{mcconnell1982spectral, morosan2018characteristics, dorovskyy2017properties, clarke2019properties}. Some of these 204 S-bursts have a clear sub-structure in the intensity and frequency drift rate. We proposed a theory to interpret the sub-structures of the Sb-bursts and the {positive} relationship between the drift rate $df/dt$ and flux intensity $I_{obs}$. With this theory, we can numerically reproduce the observed phenomena in the dynamic spectrum {as shown in Figure \ref{fig:simulation}}. We also find the existence of the multi-lane structure in some of the S-burst, {which may indicate the recurrent injection of the electron beams. For now, we are not able to give a theoretical explanation for the generation of drift pair. Future investigations concerning the drift pairs S-bursts with high time resolution radio imaging could help understand the generation process of the multi-lanes seen in S-bursts.}

According to the simulation, the wavy structure and the flux intensity variation of the Sb-burst can be re-created with the density fluctuations of background electron density at a length scale of about 1.5\,Mm. The length scale of the density fluctuations determines the length of the 'segments' in the Sb-burst sub-structures, the amplitude of $\Delta N_{e0}/N_{e0}$ determines how wavy the frequency drift line is, and the amplitude of the flux intensity variation. The simulation indicates that {$\Delta N_{e0}/N_{e0} =0.15\%$} is enough to produce the segmented fine structures. {Larger amplitude of density fluctuation may result in positive $df_{pe}/dh$. When there is positive $df_{pe}/dh$, some of the wave can be reflected back and forth (or even become trapped) in the density well. As a result, these waves experience very strong collision absorption, and the corresponding S-Burst would have some discontinuity gaps with nearly zero flux in the dynamic spectrum.} 

From the intensity distribution of the observed Sb-burst, apart from the variation between segments, we can also see that it is modulated with smaller scale and amplitude variations. These smaller scale and amplitude variation in flux intensity may result from smaller scale and amplitude density variation in the coronal electron densities. Previous remote sensing observations \citep{mugundhan2017solar} show that the $\Delta N_{e0}/N_{e0}$ varies from 0.01 to 0.1 within 1.5 to 200 solar radii. The simulation of the frequency fine structure of type III radio bursts \citep{li2012frequency} used 5\% of $\Delta N_{e0}/N_{e0} $ at length scales of 15\,Mm and produced comparable structure with the observation. There is still no density fluctuation model from in-situ observations of the inner corona ($<15$ solar radii), and the recent Parker Solar Probe mission \citep{fox2016solar} will go as close as nine solar radii, which will constrain the estimation of the amplitude and the length scale of the coronal density fluctuations.

\begin{figure}[h]
	\centering
	\includegraphics[width=0.55\linewidth]{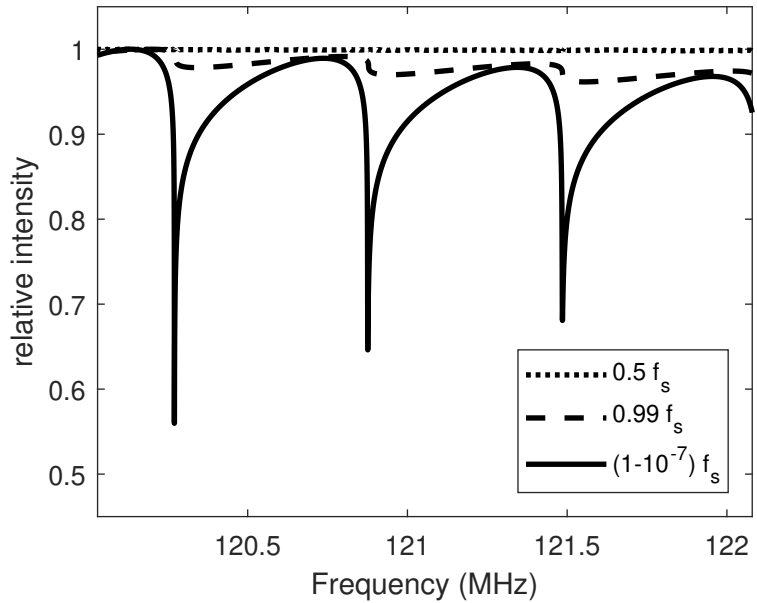}
	\caption{The flux intensity variation of the S-burst with different starting local plasma frequency}
	\label{fig:fstart}
\end{figure}

In the numerical calculation of optical depth (in Figure \ref{fig:simulation}), we assumed an observed frequency near fundamental plasma frequency for the wave generation, which means, the wave comes from regions where $f_{pe}\approx f_s$. This assumption was previously used in the numerical simulation of solar type III bursts \citep{takakura1982numerical, li2008simulations}. Previous studies \citep{mcconnell1983evidence, melnik2010solar, morosan2015lofar} also support the idea that the S-bursts waves come from a region close to the local plasma frequency. To test the sensitivity of the flux intensity variation to the starting frequency of the wave, we changed the starting frequency of the wave propagation. For the result in Figure \ref{fig:simulation}, the integration range of Equation \ref{eq:tau} is from $h_0\left|_{f_{pe0}=(1-10^{-7})f_s}\right.$ to $200$ solar radii. We considered two more cases, the starting height of the wave propagation is changed to $h_0\left|_{f_{pe0} = 0.5 f_s}\right.$ and $h_0\left|_{f_{pe0} = 0.99 f_s}\right.$, the results are shown in Figure \ref{fig:fstart}. From Figure \ref{fig:fstart}, we can see that, if the wave comes from the region of the local plasma frequency (solid line in Figure \ref{fig:fstart}), the variation in relative intensity is large. While, if the starting frequency is slightly away from the local plasma frequency, the variation of the flux intensity is smaller. \ {For the harmonic wave, the calculated intensity variation is less than $10^{-4}$. This indicates that the discontinuous fine structure is more likely to exist in the fundamental wave, not second or third harmonic ($2f_{pe}=f_s$ or  $3f_{pe}=f_s$)}. The significant intensity variation of the Sb-burst requires not only density fluctuations, but also that, the wave is generated at the place of fundamental local plasma frequency. 

{In our computation, we don't consider the temporal variation and propagation of the density fluctuations. This assumption is proposed due to the relative time scale of the fluctuation variation and the electron beam motion. The S-bursts duration is much shorter than the time scale of the background density fluctuations variation with large length scales. The speed of the electron exciter is set as $0.08\, c$, which is much larger than the Alfv{\'e}n speed in the corona. Moreover, density fluctuations can cause scattering of the radio waves during propagation, which is also not included in the simulation. The duration of the radio emission at a given frequency can also be broadened by scattering \citep{kontar2019anisotropic}. The increase of the duration may mix the segments of the discontinuity and weaken the significance of fine structures.}

Finally, we would like to note that the variation of the flux intensity due to collisional absorption is independent on the emission mechanism of S-bursts, though plasma emission is emphasized in the above discussion. If the radio wave is considered to be generated by ECM in a density-depleted magnetic flux tube \citep{wu2002generation}, the wave will be confined in the tube and propagate upwards before the local plasma frequency outside of the tube is equal to the wave frequency. If there are electron density fluctuations of the background corona outside of the magnetic tube, the variation of the flux intensity due to collisional absorption may also occur when the radio wave leaves the magnetic tube (or the apparent source region). The lower the density gradient is, the higher is the absorption, and \textit{vice versa}. Moreover, from Equation \ref{eq:speed}, the wave propagation speed is lower when the local plasma frequency is closer to the wave frequency. When the density gradient is lower, the wave will leave the apparent source region with a lower speed, and as a result, the frequency drift rate $df/dt$ is reduced. This is consistent with the above observation that the segments of S-burst with larger absolute value of frequency drift rate tend to be brighter in the dynamic spectrum. In the future, further imaging spectroscopy of the Sb-burst, especially polarization and imaging observations combined with the magnetic field topology can help interpret the emission mechanism of S-bursts.

\section{Acknowledge}
We are thankful to the ASTRON/JIVE Summer Student Programme 2019 for the financial support. This paper is based on data obtained with the International LOFAR Telescope (ILT) under project code LT10 002. LOFAR (van Haarlem et al. 2013) is the Low Frequency Array designed and constructed by ASTRON. It has observing, data processing, and data storage facilities in several countries, that are owned by various parties (each with their own funding sources), and that are collectively operated by the ILT foundation under a joint scientific policy. The ILT resources have benefitted from the following recent major funding sources: CNRS-INSU, Observatoire de Paris and Universite Orleans, France; BMBF, MIWF- NRW, MPG, Germany; Science Foundation Ireland (SFI), Department of Business, Enterprise and Innovation (DBEI), Ireland; NWO, The Netherlands; The Science and Technology Facilities Council, UK; Ministry of Science and Higher Education, Poland. We also acknowledge Clarke for providing the data point of the frequency drift rate in low frequency. The research in USTC was supported by the National Nature Science Foundation of China (41574167 and 41974199).

\appendix

\section{Qualitative analysis of Coulomb absorption}
\label{ap:1}
\ {Consider the integration of Equation \ref{eq:main} from $h_0$ where $f_{pe}=f_s$ to infinitely far away ($f_{pe}=0$)}, the optical depth can be rewritten as
\begin{align}
 \tau_c &= \int_{h_0}^{inf} \kappa_0 f_{s}^2 \frac{f_{pe}^4}{f_s^4} \frac{1}{\sqrt{1-\frac{f_{pe}^2}{f_s^2}}} dh  \nonumber \\
 &= \int_{f_s}^{0} \kappa_0 f_s^2 \frac{f_{pe}^4}{f_s^4} \frac{1}{\sqrt{1-\frac{f_{pe}^2}{f_s^2}}}\frac{1}{\frac{d f_{pe}}{dh}} df_{pe},
\label{eq:ap:1} 
\end{align}
{where $f_{pe}\leq{f_s}$ is required. For a given radio frequency $f_s$, the integrand decreases sharply with the decrease of the frequency ratio $f_{pe}/f_s$, roughly in the order of $(f_{pe}/f_s)^4$. In general, the electron density and the plasma frequency decrease with the altitude in corona, so the main contribution of $\tau_c$ comes from the plasma in a limited region near the radio source position with $f_{pe} \lesssim f_s$. As an result, the gradient $df_{pe}/dh$ can be taken out of the integration with the approximation $(df_{pe}/dh) \sim (df_{pe}/dh)|_{f_{pe}=f_s}$.} 

After applying the element changing of $u\coloneqq f_{pe}/f_s$, the integration can be simplified into \citep{gradshteyn2014table}
{
\begin{align}
 \tau_c &= \kappa_0 f_s^3 \left(-\left.{\frac{df_{pe}}{dh}}\right|_{f_{pe}=f_{s}}\right)^{-1}  \int_{0}^{1} u^4
  \frac{1}{\sqrt{1-u^2}} du  \quad 
\label{eq:ap:A2} 
  \\ 
  &=  \kappa_0 f_s^3 \left(-\left.{\frac{df_{pe}}{dh}}\right|_{f_{pe}=f_{s}}\right)^{-1}\left(\frac{3\pi}{16}\right).
\label{eq:ap:int} 
\end{align}}{The integrand has a singular point at $f_{pe}=f_s$ in  Equation \ref{eq:ap:1} or at $u=1$ in Equation \ref{eq:ap:A2}, but the integration is convergent. The optical depth is a finite value for the wave originates from the place where $f_{pe}=f_s$. From Equation \ref{eq:ap:int}, we can see that the optical depth is inversely proportional to the {local plasma frequency decreasing rate,}}
{
\begin{equation}
\tau_c \propto \left(-\left.{\frac{df_{pe}}{dh}}\right|_{f_{pe}=f_{s}}\right)^{-1}.\label{eq:tau_A}
\end{equation}
\textbf{In other words, the optical depth $\tau_c$ is determined mainly by the function $f_{pe}(h)$ near the emitting region.} This is consistent with the numerical integration result shown in Figure \ref{fig:simulation}.}

{We need to note that negative value of $df_{pe}/dh$ is assumed in above discussion. When the density fluctuation is strong enough to create a positive $df_{pe}/dh$ in the decreasing background, the wave generated at $f_{pe}=f_s$ cannot propagate toward the observer in this one-dimensional model. Consequently, the corresponding radio source is not visible to the observer.}

\section{Convergence of the Numerical Integration}

\begin{figure}[h]
	\centering
	\includegraphics[width=0.6\linewidth]{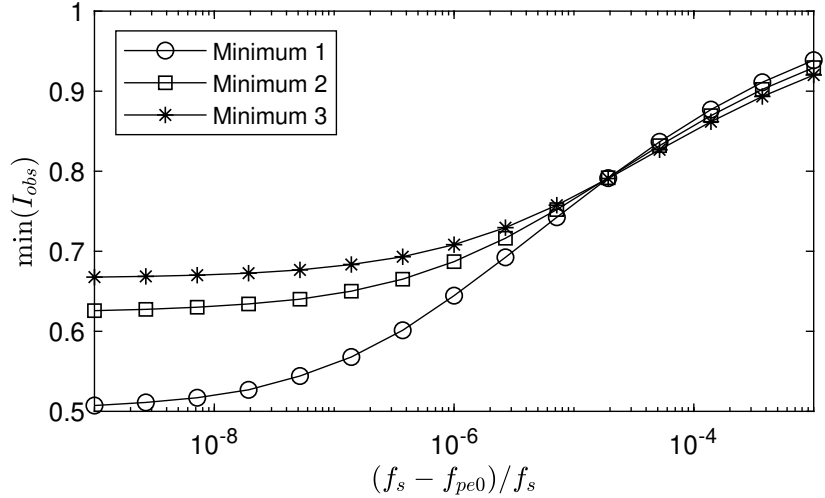}
	\caption{\textbf{The minimum value of the relative flux with different local starting frequency $f_{pe0}$. Here, the minimum 1, 2 and 3 represent the minimal points near the frequency 120.2\,MHz, 120.9\,MHz, 121.5\,MHz on the black line in the left sub-panel of Figure \ref{fig:simulation}, respectively.}}
	\label{fig:converge}
\end{figure}

\textbf{Although the integration in Equation \ref{eq:main} and \ref{eq:ap:1} is convergent theoretically, the numerical integrating result is sensitive to the starting point near the singularity as shown in  Figure \ref{fig:converge}. We can see that the minimum of the relative flux converges when the local plasma frequency of the starting point approaches the frequency of the wave as $(f_s-f_{pe0})/f_s\xrightarrow{}0$}.

\bibliography{cite}

\end{document}